# Scaling mechanism for efficient wavelength conversion in laser plasmas

Matteo Clerici[1,*], Marco Peccianti[1,2], Bruno E. Schmidt[1], Lucia Caspani[1], Mostafa Shalaby[1], Mathieu Giguère[1], Antonio Lotti[3,4], Arnaud Couairon[4], François Légaré[1], Tsuneyuki Ozaki[1], Daniele Faccio[1,5] and Roberto Morandotti[1]

[1]*INRS-EMT, 1650 Blvd. Lionel-Boulet, Varennes, Québec J3X 1S2, Canada*
[2]*Institute for Complex Systems (ISC), CNR, via dei Taurini 19, 00185 Rome, Italy*
[3]*Dipartimento di Scienza e Alta Tecnologia, Università degli Studi dell'Insubria, via Valleggio 11, 22100 Como, Italy*
[4]*Centre de Physique Théorique CNRS, Ecole Polytechnique, F-91128 Palaiseau, France*
[5]*School of Engineering and Physical Sciences, Heriot-Watt University, SUPA, Edinburgh EH14 4AS, UK*
*Corresponding author: clerici@emt.inrs.ca*

Laser-induced ionization is a fundamental tool for the frequency conversion of lasers into spectral regions so far inaccessible, including both extreme ultraviolet and terahertz. The low-frequency currents induced by laser-driven ionization generate extremely broadband, single-cycle terahertz pulses, with applications ranging from remote sensing to optical pulse diagnostic, yet strong limitations arise from the low conversion efficiencies of this mechanism. We show a remarkable increase of the radiated terahertz energy with the laser wavelength and we relate this observation to the stronger action of long-wavelength fields on ionization-induced free-carriers. Ultimately, the use of mid-infrared pulses instead of the near-infrared ones employed so far enables the unprecedented table-top generation of the extremely high terahertz fields (>4 MV/cm) required for, e.g. the optical manipulation of quantum states, attosecond pulse synthesis and time-resolved studies of ultrafast electron dynamics. Furthermore, such high fields allowed us to perform space-time resolved terahertz diagnostics exploiting standard optical components.

The coupling between optical pulses and free-carriers is extremely sensitive to the optical wavelength, and indeed the acceleration of an electron in an optical field, induced by the ponderomotive force associated to the oscillating driving field, increases with the square of the laser wavelength due to the longer optical cycle. Thanks to this mechanism, by simply switching the driving field from the standard 800 nm to longer wavelengths in the mid-infrared region ($\lambda > 1.5$ μm), it has been possible to significantly extend the cut-off frequency of the high-order harmonic generation –thus enhancing coherent soft X-ray production [1,2].

Starting from this observation we investigated the possibility to generalize this paradigm –longer wavelengths enhance the laser-electrons interactions– for improving the generation of radiation at the opposite extreme of the electromagnetic spectrum, i.e. in the terahertz (THz) frequency range, where high peak fields are not easily accessible. Currently, high-field THz pulses (peak field $\simeq 20$ MV/cm) can only be provided by large-scale, accelerator-based facilities [3], whereas table-top solutions (relying on small-scale laser systems) are still an open issue, which is mainly being addressed by optical rectification in nonlinear crystals. With this technique, however, the high –up to $10^{-2}$– conversion efficiencies from laser to THz pulses are counterbalanced by low beam quality and bandwidth, in turn resulting in peak fields currently limited to $\simeq 1.5$ MV/cm [4]. Higher field values can be achieved by difference-frequency generation in second order nonlinear crystals, yet in the multi-THz (carrier frequency >10 THz) regime only [5,6]. Therefore, a relatively simple way to further increase the THz field in the <10 THz range is now an absolute necessity for driving the study of high-field phenomena [7] as well as for boosting a very diverse range of applications [8].

Laser-induced ionization in symmetry-broken laser fields [9], obtained by focusing an optical pulse together with its second harmonic in a gas, may provide a viable solution for high-field table-top THz generation, owing to the extremely large bandwidths (>100 THz) and remarkable focusing properties of the generated pulses [10,11], that allowed to report peak field values up to 0.5 MV/cm [12]. However, until now, the attempts to further increase the THz peak field have almost exclusively relied on the brute-force approach of simply increasing the pump energy. Unfortunately this method is hampered by saturation effects due to plasma defocusing as well as other detrimental nonlinear mechanisms that act to finally decrease the THz emission, currently limited to conversion efficiencies of no more than $10^{-4}$-$10^{-5}$ [13]. No methods to significantly scale up the THz generation efficiency in plasma have been proven effective so far.

Starting from the model proposed by Kim and co-workers [13] for explaining the THz generation process in symmetry-broken laser fields, we theoretically investigated the wavelength dependence of the transverse, transient photocurrents responsible for the THz emission, with the aim of unveiling a scaling mechanism similar to that observed for high-order harmonic generation. We found that the emitted THz energy is indeed proportional to the square of the photocurrent density, which in turns scales as $\lambda$. This means that, under the assumption of a wavelength-independent excitation (plasma) volume, the THz output scales quadratically with the pump wavelength, i.e. as $\lambda^2$. This was verified experimentally by monitoring the THz radiation emitted from the electron transient photo-currents with twelve different pump laser wavelengths between 0.8 μm and 2.02 μm.

The experimental layout is sketched in Fig.1 and relies on a parabolic mirror focusing of the input two-colour laser field for the plasma and electron current excitation

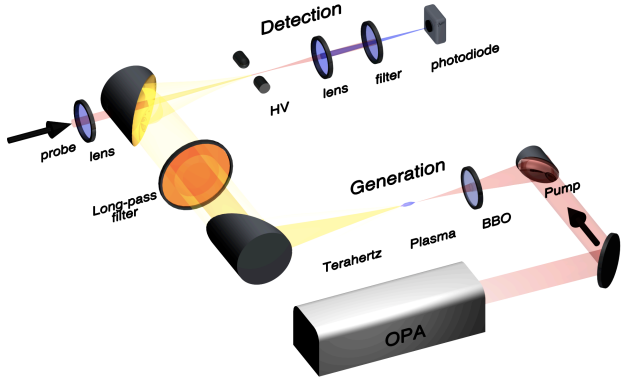

**Fig. 1** The wavelength tunable output from a commercial optical parametric amplifier is focused together with its second harmonic generated by a frequency doubling crystal. The generated THz pulse is collected by a parabolic imaging system and focused into the detection setup. The time resolved electric field is detected by way of Air Biased Coherent Detection (ABCD in figure). A pyroelectric detector and a pyroelectric camera also monitor pulse energies and beam profiles, respectively.

[14]. The total THz energy radiated from the few mm long plasma source is measured by a commercial pyroelectric detector while an Air Biased Coherent Detection scheme – ABCD– allows for recording the THz electric field in both amplitude and phase [15,16]. We tuned the input pump laser wavelength from 800 nm to 2.02 µm while maintaining all other parameters constant: 60±5 fs pulse duration, 400±5 µJ input fundamental and 20±5 µJ second harmonic energies, and input pump diameter ≈ 6 mm. The choice of the experimental parameters has been made in view of providing a direct comparison of the THz field generated at different pump wavelengths in conditions that are easily reproducible and similar to the ones commonly employed for THz generation by two-colour driven ionization. In such conditions, as the laser pump pulse is changed towards the mid-infrared spectral range, we have indeed found a remarkable increase in the emitted THz radiation energy.

Figure 2a shows the radiated THz energy as function of increasing pump wavelength. The dependence of the THz energy from the driving field wavelength is not a trivial power law, indicating a somewhat more complex phenomenology than that predicted by the model accounting only for the wavelength dependence of the driving field ponderomotive force. However, we can highlight a clear monotonic increase in the THz yield, fitted by a $\lambda^{4.6\pm0.5}$ power law, which continues up to pump wavelengths of 1.8 µm. Pump wavelengths longer than ≃ 1.8 µm show an actual decrease in the observed THz energy and conversion efficiency. In the reported data, the maximum gain in the energy conversion efficiency, calculated as the ratio between the THz energy at 1.8 µm and the one at 0.8 µm (i.e. the standard pump wavelength), reaches up to a factor thirty.

The origin of the departure from the quadratic law-power prediction may be found in various mechanisms, ranging from the actual pump beam focusing geometry to the nonlinear distortion effects on the pump pulse related for example to self-focusing, plasma-induced defocusing, ionization and nonlinear losses, all of which also depend on the wavelength (see *e.g.* Théberge *et al.* [17], while a comprehensive review is given in Couairon *et al.* [18]). A

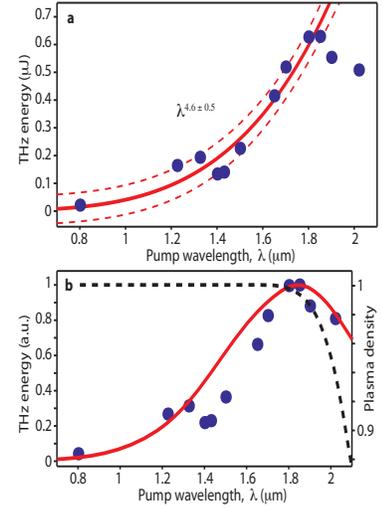

**Fig. 2** (a) Recorded THz energy for 12 different pump wavelengths, between 0.8 and 2.02 µm (solid circles). The data scale with an overall $\lambda^{4.6\pm0.5}$ trend up to wavelengths around 1.8 µm (solid red line – the dashed red lines represent the 65% confidence bounds obtained by a nonlinear least square fit). (b) Radiated THz energy dependence on the pump laser wavelength obtained by numerical integration of the transverse photocurrent model for different pump wavelengths (red-solid curve). The black dashed curve shows the overall plasma density (right axis). The energy scale in (b) is normalized and the experimental data are overlapped for clarity (solid circles).

full modelling of such nonlinear effects is beyond the scope of our report. However, we have considered a simplified model, which also accounts (in addition to the plasma current wavelength scaling) for the wavelength dependence of the beam focusing geometry, i.e. of the interaction volume (the plasma volume) and pulse intensity variations, under the assumption of linear pulse propagation. This intuitive choice is also supported by recent literature demonstrating a reduction of the nonlinear reshaping effects and of the intensity clamping for tightly focused beams, such as those used in our experiments [19,20].

The solid line in Fig. 2b depicts the results of this model, which is not only good enough to display the enhancement observed in the experiment, but also matches the inversion trend at the 1.85 µm pump wavelength. We relate this inversion to the sharp drop of the total ionization below 100% –black dashed curve in Fig. 2b– due to the increase of the focal spot size, which in turn induces a decrease in the pulse intensity.

The agreement between our model and the experiment allows us to speculate that indeed the main wavelength-scaling effects at play are related to the photocurrent amplitudes combined with linear focusing geometry effects. This hints that under the appropriate focusing conditions, e.g. for constant intensity and interaction volume, it would be possible to exploit the photo-currents amplitude dependence from the pump wavelength to further increase the THz energy for pump wavelengths longer than 2 µm.

In order to fully characterize the THz pulses generated in our experiments we recorded the electric field and beam profile. Figure 3 shows the normalized THz field temporal profiles (a) and relative power spectra (b) for an 800 nm, 1450 nm and 1850 nm pump laser pulse

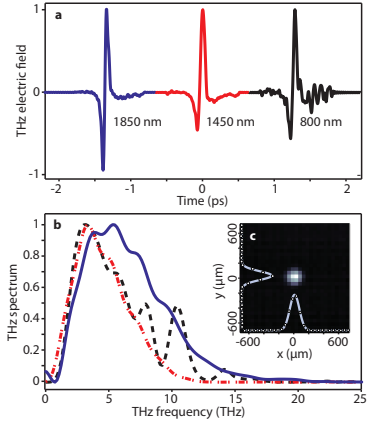

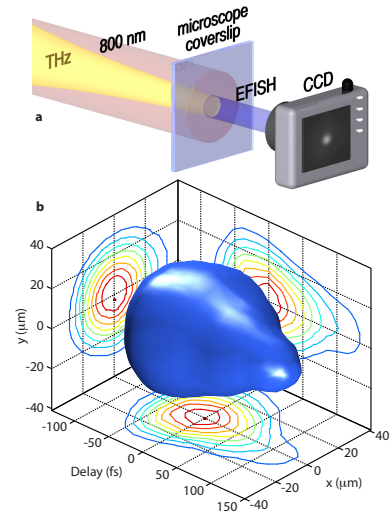

**Fig. 3** (a) Normalized THz electric field traces acquired by the ABCD detection scheme for a 400 µJ pump energy and at 1850, 1450 and 800 nm pump wavelengths (left to right), respectively. The traces are shifted in time for visualization purposes. (b) Power spectra of the THz fields in a (800nm – dashed, 1450nm – dot-dashed and 1850nm – solid). (c) THz beam profile recorded in the focus of the last parabolic mirror for the 1850nm pump wavelength. The overlay in c shows the Gaussian fit for the THz spot-size in the parabolic focus.

wavelength, respectively. The inset (c) shows the refocused THz beam profile generated by the 1850 nm pump pulse. The electric fields and spectra are relatively similar in shape confirming that we scale the THz energy whilst maintaining the temporal quality of the single cycle THz electric field pulse.

The THz electric field amplitude generated by the 1850 nm pump pulse, evaluated trough the combination of the energy (630 nJ), the beam profile and the time-resolved, un-calibrated electric field trace reaches what is, to the best of our knowledge, the highest reported value for a single cycle sub-10 THz pulse generated by a table-top source: 4.4±0.4 MV/cm, setting a completely new standard for the affordable generation of intense THz pulses, in turn paving the way to extreme THz nonlinear optics. With respect to this, our source quickly allowed us to investigate a series of intriguing phenomena, such as THz-triggered air breakdown and lasing, which will be described elsewhere. As a single, revealing example, we observed that such intense THz pulses are able to trigger the Terahertz Field Induced Second Harmonic (TFISH) [15,21] of a collimated, 800 nm optical beam in a thin slice of glass (a simple microscope coverslip). This nonlinear process in turn allows for the spatially resolved mapping of the THz beam profile with a standard, low-cost CCD camera. Furthermore, the 800 nm pulse (60 fs duration) also provides a time gate to the nonlinear interaction, rendering a full space- and time-resolved THz mapping, in a similar fashion to a well-known technique employed for the characterization of optical pulses at more standard wavelengths, namely 3D-mapping [22]. By recording the spatially resolved second harmonic signal for different THz-to-800 nm pulse delays –see Fig. 4a– we reconstructed the full, spatio-temporal THz profile, shown in Fig. 4b. We note that so far, direct beam reconstruction in this range of frequencies is typically only possible with very slow, extremely costly pyroelectric cameras (hence with no temporal information) or by electro-optical

**Fig. 4** (a) Sketch of the imaging technique relying on the TFISH process performed by the THz field (yellow) on the 800 nm beam (red), resulting in a $\lambda \simeq 400$ nm signal (blue) spatially resolved by a standard CCD camera (the picture on the camera represents a real image of the beam at the focus). (b) Recording such signal for different THz-to-800 nm pulse delays allows to reconstruct a THz 3D-map (two spatial coordinates plus the temporal one). The figure shows an iso-amplitude surface at the 20% of the maximum. The curves on the sides are iso-amplitude levels for the corresponding projections sampling in second order nonlinear media (where phase matching plays a relevant role). Further investigations are ongoing to refine this novel detection scheme enabled by the intense and broadband THz pulses resulting from mid-infrared driven ionization.

In conclusion, we show that the recently introduced paradigm of scaling to longer wavelengths in order to improve frequency up-conversion, *e.g.* to increase the cut-off frequency in high harmonic generation, applies also to down-conversion at the opposite extreme of the electromagnetic spectrum (far-infrared and terahertz). We used our scheme to generate linearly polarized THz pulses, characterized by an excellent near wavelength-limited focusability –see inset in Fig. 3c. More noticeably, our novel method delivers energy conversion efficiencies from the mid-infrared pump pulse to the THz pulse larger than $10^{-3}$, i.e. almost two orders of magnitude larger than in previous plasma-based generation experiments performed with an 800 nm pump wavelength at similar laser pump energies. Among the intriguing applications of this new source, we demonstrated a full spatio-temporal 3D-mapping of the THz pulses, obtained using a simple microscope coverslip and a standard (low-cost) CCD camera.

### References

1. J. Tate, T. Auguste, H. Muller, P. Salières, P. Agostini, and L. DiMauro, Phys. Rev. Lett. **98**, 013901 (2007).
2. P. Colosimo, G. Doumy, C. I. Blaga, J. Wheeler, C. Hauri, F. Catoire, J. Tate, R. Chirla, A. M. March, G. G. Paulus, H. G. Muller, P. Agostini, and L. F. DiMauro, Nature Phys. **4**, 386 (2008).
3. D. Daranciang, J. Goodfellow, M. Fuchs, H. Wen, S. Ghimire, D. A. Reis, H. Loos, A. S. Fisher, and A. M. Lindenberg, Appl. Phys. Lett. **99**, 141117 (2011).
4. C. P. Hauri, C. Ruchert, C. Vicario, and F. Ardana, Appl.

Phys. Lett. **99**, 161116 (2011).
5. A. Sell, A. Leitenstorfer, and R. Huber, Opt. Lett. **33**, 2767 (2008).
6. F. Junginger, A. Sell, O. Schubert, B. Mayer, D. Brida, M. Marangoni, G. Cerullo, A. Leitenstorfer, and R. Huber, Opt. Lett. **35**, 2645 (2010).
7. J. A. Fülöp, L. Pálfalvi, M. C. Hoffmann, and J. Hebling, Opt. Express **19**, 15090 (2011).
8. M. Tonouchi, Nature Photon. **1**, 97 (2007).
9. D. Cook and R. Hochstrasser, Opt. Lett. **25**, 1210 (2000).
10. K. Y. Kim, a. J. Taylor, J. H. Glownia, and G. Rodriguez, Nature Photon. **2**, 605 (2008).
11. M. D. Thomson, V. Blank, and H. G. Roskos, Opt. Express **18**, 23173 (2010).
12. T. Bartel, P. Gaal, K. Reimann, M. Woerner, and T. Elsaesser, Opt. Lett. **30**, 2805 (2005).
13. K.-Y. Kim, J. H. Glownia, A. J. Taylor, and G. Rodriguez, Opt. Express **15**, 4577 (2007).
14. F. Blanchard, G. Sharma, X. Ropagnol, L. Razzari, R. Morandotti, and T. Ozaki, Opt. Express **17**, 6044 (2009).
15. J. Dai, X. Xie, and X. C. Zhang, Phys. Rev. Lett. **97**, 103903 (2006).
16. N. Karpowicz, J. Dai, X. Lu, Y. Chen, M. Yamaguchi, H. Zhao, X. C. Zhang, L. Zhang, C. Zhang, M. Price-Gallagher, C. Fletcher, O. Mamer, A. Lesimple, and K. Johnson, Appl. Phys. Lett. **92**, 011131 (2008).
17. F. Théberge, W. Liu, P. Simard, A. Becker, and S. Chin, Phys. Rev. E **74**, 036406 (2006).
18. A. Couairon and A. Mysyrowicz, Physics Reports **441**, 47 (2007).
19. M. Gaarde and A. Couairon, Phys. Rev. Lett. **103**, 043901 (2009).
20. P. Prem Kiran, S. Bagchi, C. Arnold, S. R. Krishnan, G. R. Kumar, and A. Couairon, Opt. Express **18**, 21504 (2010).
21. A. Nahata and T. F. Heinz, Opt. Lett. **23**, 67 (1998).
22. M. A. C. Potenza, S. Minardi, J. Trull, G. Blasi, D. Salerno, A. Varanavičius, A. Piskarskas, and P. Di Trapani, Opt. Comm. **229**, 381 (2004).